\documentclass{aastex631}

\usepackage{xspace} 
\newcommand{\eos}{\textsc{Eos}\xspace}
\newcommand{\edataset}{\textsc{Dataset}\xspace}

\def\d{{\mathrm{d}}}

\newcommand{\es}{\end{subequations}}

\newcommand{\be}{\begin{equation}}
\newcommand{\ee}{\end{equation}}
\renewcommand{\d}{{\rm d}}

\newcommand{\llp}{\left [}
\newcommand{\rrp}{\right ]}
\newcommand{\lp}{\left (}
\newcommand{\rp}{\right )}

\def\lsim{\mathrel{\rlap{\lower4pt\hbox{\hskip0.5pt$\sim$}}
    \raise1pt\hbox{$<$}}}         %less than or approx. symbol
\def\gsim{\mathrel{\rlap{\lower4pt\hbox{\hskip0.5pt$\sim$}}
    \raise1pt\hbox{$>$}}}         %greater than or approx. symbol

\usepackage{nicematrix}

\renewcommand{\NG}{\text{\tiny NG}}
\newcommand{\fNL}{f_\text{\tiny NL}}

\newcommand{\sapienza}{Dipartimento di Fisica, Sapienza Università 
	di Roma, Piazzale Aldo Moro 5, 00185, Roma, Italy}
\newcommand{\infn}{INFN, Sezione di Roma, Piazzale Aldo Moro 2, 00185, Roma, Italy}

\newcommand{\unige}{D\'epartement de Physique Th\'eorique and Centre for Astroparticle Physics (CAP), Universit\'e de Gen\`eve, 24 quai E. Ansermet, CH-1211 Geneva, Switzerland}

\begin{document}

\title{
High-redshift JWST Observations and Primordial Non-Gaussianity
}

\author{Matteo Biagetti}
% \email{}
\address{Institute for Fundamental Physics of the Universe, Via Beirut 2, 34151 Trieste, Italy}
\address{SISSA - International School for Advanced Studies, Via Bonomea 265, 34136 Trieste, Italy}
\address{Istituto Nazionale di Astrofisica, Osservatorio Astronomico di Trieste, via Tiepolo 11, 34143 Trieste, Italy}
\address{Istituto Nazionale di Fisica Nucleare, Sezione di Trieste, via Valerio 2, 34127 Trieste, Italy}

\author{Gabriele Franciolini}
% \email{gabriele.franciolini@uniroma1.it}
\affiliation{\sapienza}
\affiliation{\infn}

\author{Antonio Riotto}
% \email{Antonio.Riotto@unige.ch}
\affiliation{\unige}
\affiliation{Gravitational Wave Science Center (GWSC), Universit\'e de Gen\`eve, CH-1211 Geneva, Switzerland}

\date{\today}

%\maketitle

\begin{abstract}
\noindent
Several bright and massive galaxy candidates at high redshifts have been recently observed by the James Webb Space Telescope. 
Such early massive galaxies seem difficult  to reconcile with standard $\Lambda$ Cold Dark Matter model predictions. 
We discuss under which circumstances such observed massive galaxy candidates can be explained by introducing primordial non-Gaussianity in the initial conditions of the cosmological perturbations.
\end{abstract}

\section{Introduction}\label{intro}
\noindent
The standard cosmological model, based on the idea that the energy budget of the universe is currently dominated by a tiny cosmological constant $\Lambda$ plus 
Cold Dark Matter ($\Lambda$CDM), predicts that the initial seeds for galaxy formation are  halos with relatively low masses of the order of $10^6 M_\odot$.

The initial James Webb Space Telescope (JWST) imaging 
via early release programs, such as Cosmic Evolution Early Release Science (CEERS), early release observations (ERO) and early release science (ERS),
% the Cosmic Evolution Early Release Science (CEERS) survey 
has recently reported  a population of surprisingly massive galaxy candidates at redshift $z\gsim 8$ with stellar masses of the order of $10^9 M_\odot$ \cite{2022arXiv220712338A,2022arXiv220712474F,2022arXiv220801612H,2022arXiv220802794N,Yan:2022sxd}.
Even though a spectroscopic follow-up will be necessary to confirm the observation based on  photometry only, the early formation of massive galaxies reported by the JWST is hardly   reconcilable with the standard $\Lambda$CDM expectations, which would require  an implausible   high star formation efficiency (SFE), even larger  than  the cosmic baryon mass budget in collapsed structures. 
It is important to stress though that various uncertainties 
affects the JWST measurements and  might solve the tension 
with $\Lambda$CDM. 
For example, the calibration of JWST data may cause imprecise redshfit determination (see e.g. \cite{2023MNRAS.518.4755A,2022arXiv220807879S}), while the estimation of the stellar masses may be plagued by 
systematic uncertainties on the initial mass distribution,
the effect of a large scatter in the star formation  \cite{2022arXiv220812826M}, the impact of dust attenuation \cite{2022arXiv220906840Z}, as well as the inclusion of very bright lines from other sources beyond the stellar continuum (e.g. \cite{2022arXiv220814999E}).
 The spectroscopic follow-up and further testing on the astrophysical uncertainties will soon shed more light on the issue.

A useful quantity to assess the viability  of the $\Lambda$CDM model is  the stellar mass density $\rho_*(>M_*)$ 
predicted above a given mass scale $M_*$.  
The stellar mass is related to the 
average baryon mass within each halo through 
the SFE, which we define as $\epsilon$, by the relation
\begin{equation}
    M_* = \epsilon 
    (\Omega_\text{\tiny b}/\Omega_\text{\tiny m}) M
    = \epsilon  f_\text{\tiny b} M,
    \end{equation}
    with $M$ being the halo mass and $f_\text{\tiny b} = 0.156$ the baryon fraction as measured by Planck \cite{Planck:2018vyg}.
In the following, and in  order to be on the conservative side, we will 
identify the stellar mass with the baryon mass contained within a given halo, that means fixing the SFE to $\epsilon =1$. This conservative choice maximises the stellar mass predicted by a given scenario.

The comoving cumulative stellar mass density 
contained within galaxies above a certain stellar mass $M_{\star}$ reads
\begin{align}
    \rho_*(>M_*,z)
    &
    =\epsilon f_{\rm b}
    \int_{M_* / (\epsilon f_\text{\tiny b})}^{\infty}
    \frac{\d n(M,z)}{\d M} M \d M\ , \label{esmd}
\end{align}
where $n(M)$ is the CDM halo mass function.

Recently, based on 14 galaxy candidates with masses in the range $\sim 10^{9}\div 
10^{11}\ M_{\odot}$ at $7<z<11$ identified in the JWST CEERS program,   Ref.~\cite{2022arXiv220712446L}  derived the cumulative stellar mass density at $z=8$ and 10 for $M_{\star}\gtrsim 10^{10}\ M_{\odot}$. They found
at $z\simeq 10$
\begin{align}
&\rho_*(>10^{10} M_\odot)\simeq 1.3^{+1.1}_{-0.6}\cdot 10^6 M_\odot\,{\rm Mpc}^{-3},
\nonumber\\
&\rho_*(>10^{10.5} M_\odot)\simeq 9^{+11}_{-6}\cdot 10^5 M_\odot\,{\rm Mpc}^{-3}.
\end{align}
These values are larger than the $\Lambda$CDM predictions by a factor $\sim 50$, even allowing maximum efficiency $\epsilon=1$, or invoking extreme value statistics \cite{Lovell:2022bhx}.

While several extensions of the $\Lambda$CDM scenario have been already put forward in the recent literature
\cite{Menci:2022wia,Liu:2022bvr,Gong:2022qjx}, they all appeal to new ingredients in the late time evolution of the universe. The goal of this paper is to discuss a possible solution which invokes a change in the initial conditions of the cosmological perturbations giving rise to the DM halos, that is, non-Gaussianity (NG) \cite{Bartolo:2004if}.  Indeed, a possible source of NG could be primordial in origin, being specific to a particular mechanism for the generation of the cosmological perturbations. It is known that NG in the initial conditions may change  the abundance of DM halos, especially in the high mass range of the halo mass function. As such, primordial NG may provide in principle a boost in forming high massive and bright galaxies. In the following, we characterize the nature of NG, specifying which properties NG has to possess to be in agreement with the JWST data.

The paper is organized as follows. In Sec.~II we discuss how one can model the Gaussian and NG halo mass functions, by also checking their validity with dedicated N-body simulations. 
In Sec.~III we compare models with various NG signatures to the JWST data, while our conclusions are offered in Sec.~IV.

\section{Halo mass function}

\subsection{Gaussian}
\noindent
We describe the Gaussian differential halo abundance as
\begin{equation}
    {\d n\over \d M} = 
    % \mc{R}\times 
    F(\nu){\overline{\rho}_\text{\tiny M}\over M^2} 
    {\d \ln\sigma^{-1}\over \d \ln M},
    \label{dndm}
\end{equation} 
where $\overline{\rho}_\text{\tiny M}$
is the background average matter density, $\nu=\delta_c/\sigma(M,z)$ with 
$\delta_c = 1.686$ corresponding to the critical linear overdensity for collapse, while
$\sigma(M ,z)$ being the variance of the smoothed linear density field.
The smoothing scale $R$ is related to the halo mass through the relation
$R=\lp {3M}/{4\pi\overline{\rho}_\text{\tiny M}}\rp^{1/3}$.    
Linear density fields evolve with time according to the linear growth factor $D(z)$ and we assume a CDM form for the linear power spectrum. 
The variance of linear density perturbations smoothed on scale $R$ is therefore computed as 
\be 
\sigma^2
=\langle \delta^2 \rangle
=
\int
{\d^3 k\over (2 \pi)^3}  
W^2(kR)
\mathcal M^2(k,z)
P_\zeta(k)
\label{vari},
\ee  
where $P_\zeta(k)$ is the linear comoving curvature power spectrum, defined from the curvature field two-point function 
\begin{equation}
    \langle \zeta (\vec k_1) \zeta (\vec  k_2)\rangle \equiv \lp 2 \pi\rp^{3}
    \delta_\text{\tiny D} (\vec k_1 +\vec k_2) P_\zeta(k_1).
\end{equation}
Also, we introduced the Fourier transform of a top-hat spherical window function
\be 
W(kR)=3 \lp {{\sin(kR)\over (kR)^3}-{\cos(kR)\over (kR)^2}}\rp , 
\ee
and we defined 
\begin{equation}
    \mathcal M(k,z) = \frac{2}{5} \frac{k^2 T(k)D(z)}{\Omega_\text{\tiny M} H_0^2},
\end{equation}
in terms of the linear transfer function $T(k)$, the matter abundance $\Omega_\text{\tiny M}$ and present day Hubble rate $H_0,$
following the standard conventions in the literature.

\subsection{Non-Gaussianity}
\noindent

The presence of NGs in the initial conditions alters the  abundance of dark matter halos. Several ways of modeling this effect have been proposed in the past (see e.g.  \cite{Biagetti:2019bnp} for a recent review, and references therein). 

The general approach is based on the Edgeworth expansions of the Probability Distribution Function (PDF) of the matter density field, or of the level excursion probability of overcoming a threshold for collapse \cite{Matarrese:2000iz,LoVerde:2007ri,Desjacques:2009jb}. In the limit of weak enough NG, the expansion is usually truncated to the leading term, which is generated by the three-point function of the primordial field. 
As a result, the exponential tail of the mass function \eqref{dndm} is modified by a non-vanishing skewness and one can correct the Gaussian halo mass function with a multiplicative 
factor,
\begin{equation}
    {\d n_\NG \over \d M}= 
    {\d n\over \d M} 
    \times C_\NG(M),
\end{equation}
which we take to be the one proposed by \cite{Desjacques:2009jb},
\begin{equation} 
C_\text{\tiny NG}(M) =  
\left[
\frac{\hat \delta_c^2}{6 \Delta} \frac{\d S_3}{\d \ln \sigma} + \Delta 
\right]
\times
\exp \left( \frac{S_3 \hat \delta_c^3}{ 6 \sigma^2} \right).
\end{equation}
Here we introduced 
$\hat \delta_c = 0.949 \times \delta_c $, $\Delta \equiv \sqrt{ 1 - \hat \delta_c S_3 / 3}$  and 
the skewness $S_3$ can be computed by integrating the matter bispectrum
\begin{align}\label{S3skew}
S_3  
\equiv
\frac{\langle \delta^3\rangle }{\sigma^4 } 
=
\frac{1}{\sigma^4}
\int
\lp 
\prod_{i=1}^3
{\d^3 k_i\over (2 \pi)^3}  
\rp
\mathcal{B}(\mathbf{k}_1,\mathbf{k}_2,\mathbf{k}_3),
\end{align}
which in turn is sourced by the primordial curvature bispectrum $B_\zeta$ through 
\begin{equation}
\mathcal{B}(\mathbf{k}_1,\mathbf{k}_2,\mathbf{k}_3)=
\mathcal M(k_1,z)
\mathcal M(k_2,z)
\mathcal M(k_3,z)
\times B_\zeta(k_1,k_2,k_3).
% \langle \zeta(k_1)\zeta(k_2)\zeta(k_3)\rangle.
\end{equation}
The specific type of NG that sources the change in the halo mass function is fully specified by $B_\zeta$ in a model-dependent way. In this work, we focus on the so-called \emph{local}-type NG, which include the class of models where local interactions among fields take place on superhorizon scales (see \cite{Bartolo:2004if} for a review).

For these models, the primordial bispectrum takes the simple, factorizable, form of
\begin{equation}
B_\zeta(k_1, k_2, k_3) = \frac{6}{5} \fNL [ P_\zeta(k_2)P_\zeta(k_3) + {\rm perm}],
\label{eq:local}
\end{equation}
where $\fNL$ parametrizes the amplitude of the bispectrum and $P_\zeta$ is the primordial curvature power spectrum.\footnote{Assuming a constant $\fNL$, one can show by directly integrating Eq.~\eqref{S3skew} that an accurate fit of the skewness as a function of both scale and redshift is given by (see e.g. \cite{Chongchitnan:2010xz})
 \begin{equation}
 S_3 (M,z)= 
 \frac{1.8\times 10^{-4} \fNL }{\sigma^{0.838}(M,z)D^{0.162}(z)},
\end{equation} 
that we adopt in the remainder of this work when dealing with a constant $\fNL$. We have checked that the fit is accurate even up to redshifts $z \simeq 10$.}
While in the most popular version of the local NG $\fNL$ is scale-independent, in our comparison with JWST data we are going to test extensions that allow $\fNL$ to run with scale.
This generalization is well-motivated for several models of interactions taking place during inflation \cite{Chen:2005fe,Khoury:2008wj,Byrnes:2010ft,Riotto:2010nh,Huang:2010cy,Huang:2010es,Byrnes:2011gh} and its implications have been thoroughly investigated in CMB observations and for galaxy surveys observing at low redshift \cite{LoVerde:2007ri,Sefusatti:2009xu,Becker:2010hx,Giannantonio:2011ya,Becker:2012yr,Agullo:2012cs,Biagetti:2013sr}.
The corresponding bispectrum in this scale dependent 
model is
\begin{equation}
B(k_1, k_2, k_3) = \frac{6}{5} [\fNL(k_1) P_\zeta(k_2)P_\zeta(k_3) + {\rm perm}],
\label{eq:fnlk_bispec}
\end{equation}
where different functional forms for $\fNL(k)$ we adopt are specified in the next section.

\subsection{Testing high-redshift halo mass functions with N-body simulations}
\noindent
Previous literature has thoroughly compared theoretical predictions of the halo mass function both for Gaussian \cite{Jenkins:2000bv,VIRGO:2001szp,Warren:2005ey,Reed:2003sq,Reed:2006rw,Lukic:2007fc,Cohn:2007xu,Tinker:2008ff,Tinker:2010my,Despali:2015yla} and NG \cite{Moscardini:1990zh,Weinberg:1991qe,Matarrese:1991sj,Park:1991mh,Gooding:1991ys,Borgani:1993nz,Dalal:2007cu,Pillepich:2008ka,Achitouv:2011sq,Achitouv:2013oea,Stahl:2022did} initial conditions to simulations. However, most results are at low redshifts, $z \lesssim 2$, and none of them include comparisons at higher redshift including NG initial conditions. Hence, it is important to validate the predictions at redshifts of relevance for the galaxies observed by JWST that are discussed in this paper.

To perform our analysis, we use a subset of the \eos \edataset\footnote{Information on the \eos suite is available at \url{https://mbiagetti.gitlab.io/cosmos/nbody/eos/}},
that includes simulations with Gaussian as well as NG initial conditions. The initial particle displacement is generated at $z_{in}=99$ using \texttt{2LPTic} \cite{Scoccimarro:1997gr},  and its extended version \cite{Scoccimarro:2011pz} for local NG initial conditions, using $f_{\rm NL}=500$ as value for the non-linearity parameter. The linear power spectrum given as an input is computed using CLASS \cite{Blas:2011rf} and assumes a flat $\Lambda$CDM cosmology with $n_s=0.967$, $\sigma_8=0.85$, $h=0.7$, $\Omega_m=0.3$ and $\Omega_b=0.045$. The public code \texttt{Gadget2} \cite{Springel:2005mi} is used to evolve $512^3$ particles in a cubic box of $64$ Mpc$/h$ per side, which allows enough resolution to resolve dark matter halos down to $M \sim 10^{10} M_\odot$. We run $30$ different realizations for both the Gaussian and NG initial conditions. 
We identify halos in each simulation using the code \texttt{Rockstar} \cite{Behroozi:2011ju}, with a lower mass cut off of a minimum of $100$ particles, resulting in halos of minimum $M_{\rm min} \simeq 2.3\times 10^{10} M_\odot$. The algorithm used is Friends-of-Friends (FoF) with a
linking length $\lambda = 0.28$ at redshifts $z=8$ and $10$ and it estimates the halo mass with a Spherical Overdensity approach, with overdensity $\Delta = 200\, \bar \rho_{\rm \tiny M}$.

As already shown in \cite{Biagetti:2016ywx} on a similar set of halos at redshifts $z=0$, $1$ and $2$, the Tinker fit \cite{Tinker:2010my} provides a good agreement with the halo mass function measured on the simulations. Hence, in what follows, we adopt the Tinker halo mass function parametrized as
\begin{align}
 & F_\text{\tiny Tinker} = 0.368
 \llp
 1+
 \lp \beta\nu \rp ^{-2\phi}
 \rrp
 \nu^{2\eta+1}e^{-\gamma\nu^2/2},
 \nonumber \\
& \qquad \beta = 0.589(1+z)^{0.2}, \phi=-0.729(1+z)^{-0.08},
\nonumber\\
& \qquad \eta = -0.243(1+z)^{0.27}, \gamma = 0.864(1+z)^{-0.01},
\end{align}
where $\nu$ is the computing using the same linear power spectrum provided as input to the simulations.

\begin{figure}[t!]
	\centering
	\includegraphics[width=0.5\textwidth]{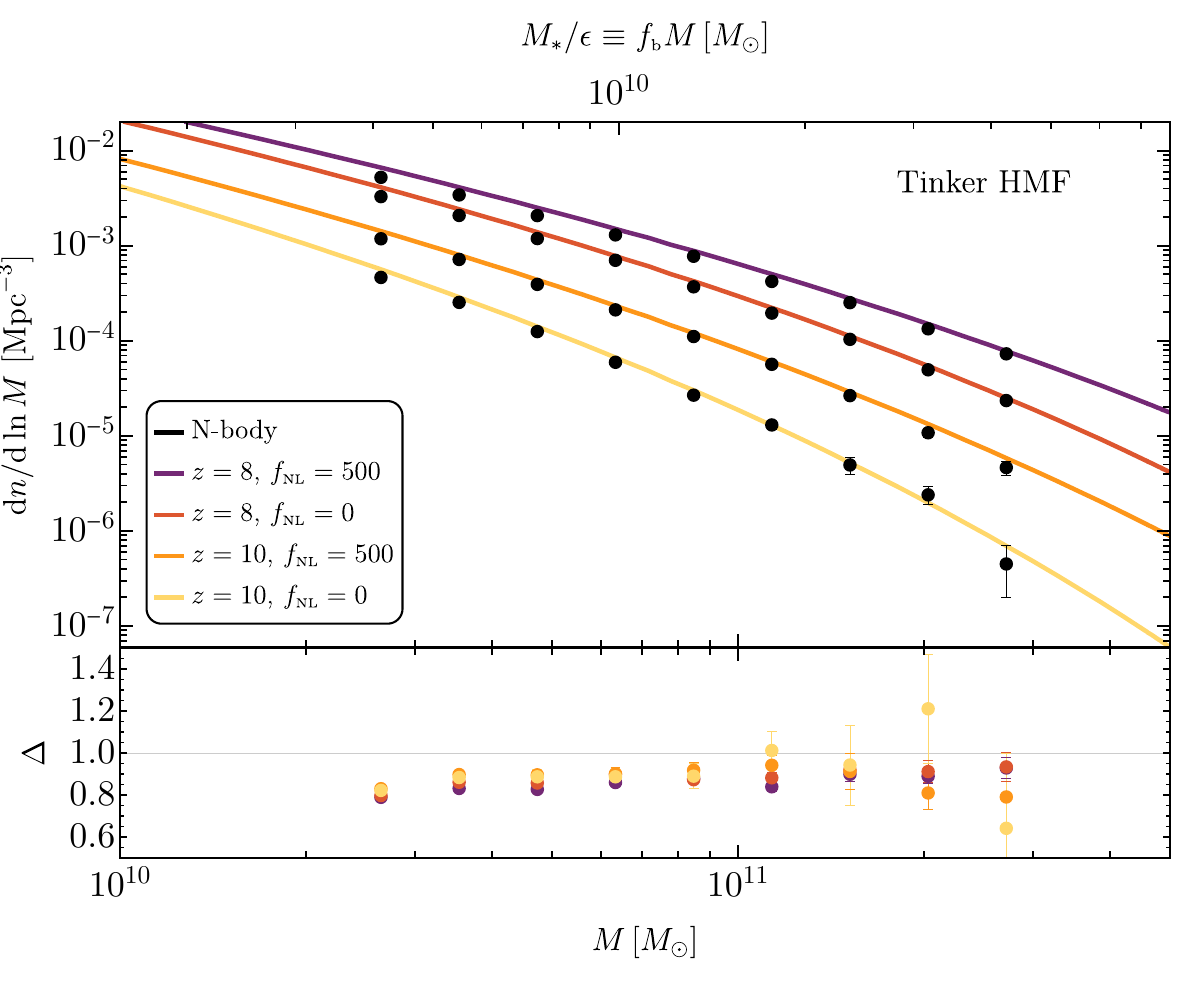}
	\caption{
	Halo mass distribution at redshift $z = 8$ and $z = 10$
	assuming either Gaussian or non-Gaussian ($\fNL = 500$) curvature perturbations and compared to N-body simulations (see text).
	The bands around the simulation data points indicate the standard error on the mean.
	 In the bottom panel, we show the ratio between the simulation data-points and the Tinker fit 
$\Delta \equiv  
\lp \d n_\text{\tiny sim}/\d \ln M \rp/
\lp \d n_\text{\tiny Tinker}/\d \ln M \rp$, adopting the same color code as in the top panel. 
}
	\label{fig:dndm}
\end{figure}

In Fig.~\ref{fig:dndm}, we show the halo mass function at various redshifts in the absence on NGs  and assuming NG initial conditions of the local type with $f_{\rm NL}=500$.  The model agrees within $20\%$ of the measurements both for Gaussian and NG initial conditions even at $z=8$ and $10$. These differences are reasonable, given that the Tinker mass function has not been tested at such high redshift and that the Rockstar halo finder is not a fully SO (Spherical Overdensity) algorithm. Thus, we are confident that our theoretical predictions for the HMF are realistic within the approximations made.

\section{The JWST data and non-Gaussianities}\label{sec:JWSTandNGs}

\noindent
Based on the results of the previous section, 
we compute the (co-moving) cumulative stellar mass density contained within galaxies above a certain stellar mass $M_{\star}$ integrating Eq.~(\ref{esmd})
including the presence of local NG. For these computations we use a value of $\sigma_8=0.815$, which is closer to the current best-fit model quoted in the latest Planck results \cite{Planck:2018vyg}. All other cosmological parameters are taken to be the same as the simulated data presented in the previous section.

\begin{figure}[t!]
	\centering
	\includegraphics[width=0.49\textwidth]{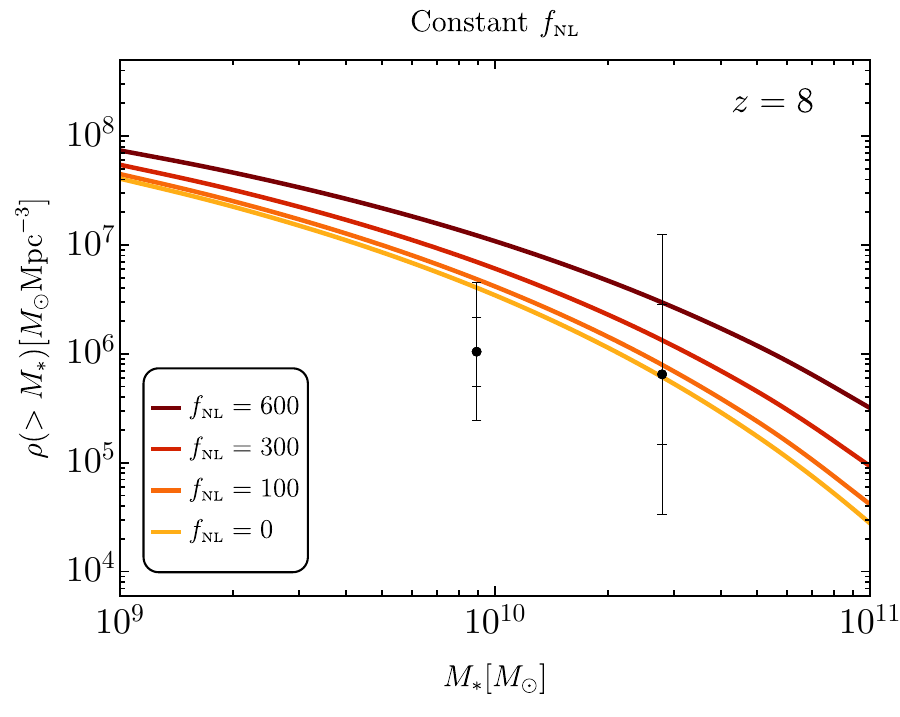}
	\includegraphics[width=0.49\textwidth]{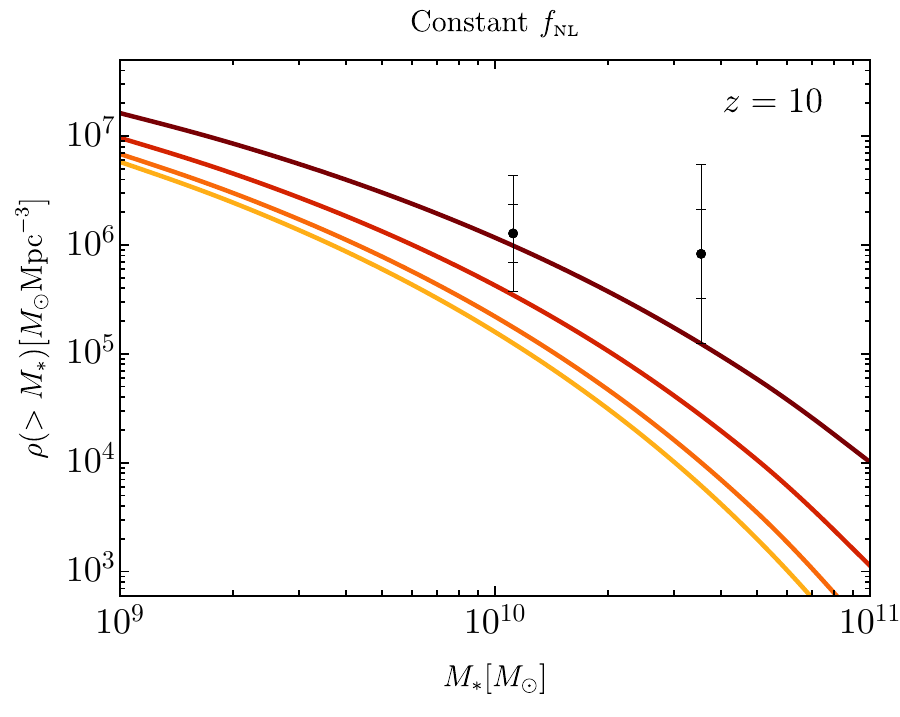}
	\caption{{\textit{Left:}}
	Co-moving cumulative stellar mass density within galaxies with
	stellar mass above $M_*$ 
	at redshift $z = 10$.
	The black bars indicate 1 and 2 $\sigma$ range inferred from the JWST observations \cite{2022arXiv220712446L}, where the latter is extrapolated assuming a Gaussian distribution.
	The same convention is used in the following figures.
	For a comparison of the JWST observations of \cite{2022arXiv220712446L} with other datasets, see for example \cite{Lovell:2022bhx}.
	{\textit{Right:}}
	Same as top for $z = 8$.
	}
	\label{fig:Labbe}
\end{figure}

In Fig.~\ref{fig:Labbe} we show the comparison between the JWST observations from Ref.~\cite{2022arXiv220712446L} and the heavy halo star density for different values of $\fNL$ in the case in which $\fNL$ is constant.
Large NGs can easily reduce the tension with observations at redshift $z \approx 10$
but do not help explaining the mild evolution between the two redshift bins. 

Such large NG are however ruled out by CMB anisotropy data \cite{Planck:2019kim} and eBOSS clustering data \cite{Castorina:2019wmr}, which constrain local-type NG to be of order $|\fNL| \lesssim 10$ and $|\fNL| \lesssim 26$ at $95\%$ confidence level, respectively.
On the other hand, one should take into account the fact that  these constraints are valid at relatively large scales, $k_{\rm constraints} \lesssim 0.3\,h/$Mpc, while the relevant scale for these massive galaxies at redshifts $z=8$ and $10$ is $k \sim 1/R \gtrsim 1.5\,h/$Mpc, where we choose $R$ to be the Lagrangian radius corresponding to halo masses of $M\sim 10^{11} M_\odot$ considered in our analysis (i.e. stellar masses around $M_*\sim10^{10} M_\odot$). Around these small scales, Ref.~\cite{Sabti:2020ser} have put constraints using UV galaxy luminosity functions from the Hubble Space Telescope (HST) of about $\fNL \lesssim 500$ at $95\%$ confidence level, but assuming that NG are switched on 
already at scales $k_{\rm cut} \sim 0.15 \, h/$Mpc. With increasing $k_{\rm cut}$, the constraints  loosen considerably \cite{Sabti:2020ser}.

We therefore  consider the possibility that the $\fNL$ parameter runs enough with scale to evade current constraints, and simultaneously explain the JWST high-redshift galaxies.

\begin{figure*}[t!]
	\centering
\includegraphics[width=0.99\textwidth]{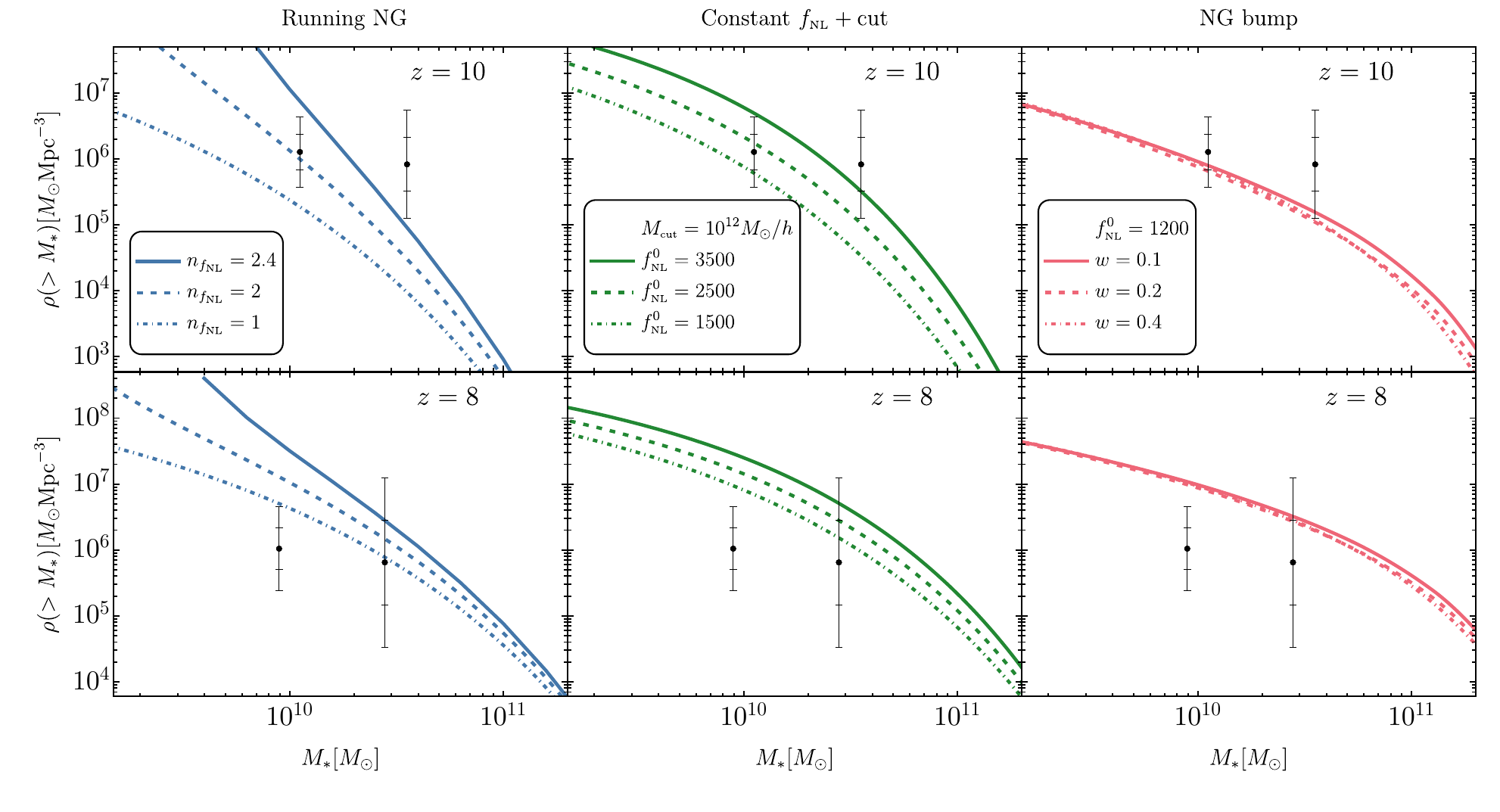}
\caption{
Stellar mass density above $M_*$ 
as shown in Fig.~\ref{fig:Labbe}
assuming different NG models.
 We emphasize that we conservatively 
assume the stellar mass is comparable to the baryon mass contained
within a given halo ($\epsilon \simeq 1$).
 As such,
% As realistic values require $\epsilon <1$, 
a satisfactory resolution of the tension would require lines to fall above the data points.
Note that the corresponding value of $f_\text{\tiny NL}^0$ should depend on $\epsilon$ if more realistic values ($\epsilon < 1$) are chosen.
{\textit{Left:}}
 The running-NG model.
\textit{Center:}
Constant NG with a sharp cut 
at the scale corresponding to halo masses $M_\text{\tiny cut}$. 
In the lower panel, transitions to negligible values of $S_3$ brings the predictions towards the Gaussian case (as can be seen in the lower panel).
{\textit{Right:}}
NG bump.}
	\label{fig:Labberunning}
\end{figure*}

\begin{figure}[t!]
	\centering
	\includegraphics[width=.99\textwidth]{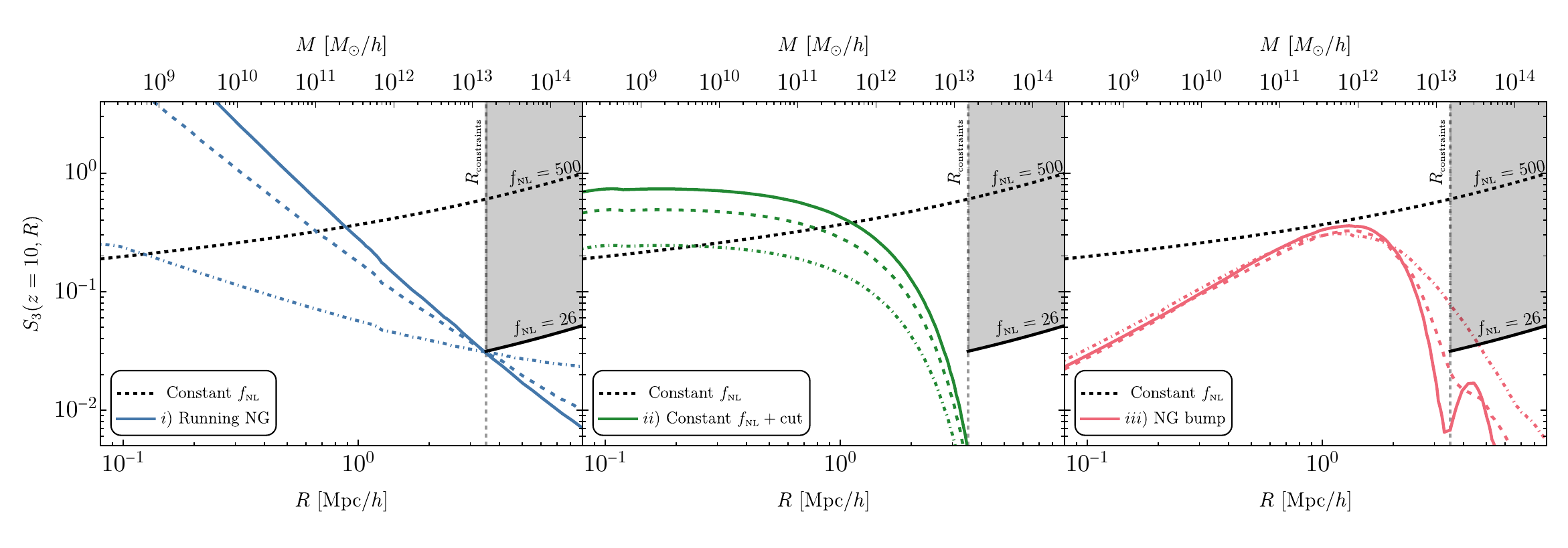}
	\caption{Skewness $S_3$ as a function of scale $R$ (or halo mass at $z=10$ indicated on top).
	The gray region corresponds to values of $S_3$ obtained using excluded values of $\fNL$ due to constraints from \cite{Castorina:2019wmr}, assuming local-type NG. We use the limiting value of $\fNL=26$ at $95\%$ confidence level using a $k_{\rm max} = 0.3\,h/$Mpc for a conservative assumption on the response of quasars to NG.
  We indicate $S_3$ obtained with the various models considered in this work with the same color used in Fig.~\ref{fig:Labberunning} and labelled in the inset. }
	\label{fig:S3running}
\end{figure}

A first case is the so-called running NG \cite{Sefusatti:2009xu} for which
\begin{equation}\label{eq:runNG}
{\it i)}\qquad
\fNL(k) = \fNL^0 
\lp \frac{k}{k_\text{\tiny max}}\rp ^{n_{\fNL}},
\end{equation}
where, depending on the running, we consider
\begin{align}
    \fNL^0 = 17 \quad \text{for}& \quad n_{\fNL} = 1,
    \nonumber \\
    \fNL^0 = 8.3 \quad \text{for}& \quad n_{\fNL} = 2,
    \nonumber \\
    \fNL^0 = 5.6 \quad \text{for}& \quad n_{\fNL} = 2.4.
\end{align}
The corresponding stellar mass densities are plotted in Fig.~\ref{fig:Labberunning} (left panel) and compared to the JWST data.
We observe that a sufficiently large $n_{\fNL} \gsim 2$ 
may help in reducing the tension,
but give rise to a too steep halo mass function tilted towards small halo masses that can hardly reach the largest datapoint at redshift $z = 10$, while being compatible with the others.
In Fig.~\ref{fig:S3running} we plot the corresponding skewness $S_3$, where we have chosen 
$k_\text{\tiny max} \simeq k_\text{\tiny constraints}$ 
to be the smallest scale constrained by LSS observations (see Ref.~\cite{Sabti:2020ser}). 
The amplitude of $\fNL(k_\text{\tiny max})$ has been fixed 
such that $S_3$ saturates the current bound from the LSS.\footnote{Note that the model of Eq. \eqref{eq:runNG}, besides being too step to explain JWST observations, produces also a very large skewness at small scales (cfr. Fig. \ref{fig:S3running}). These large values would be anyway in contrast with the truncation made on the Edgeworth expansion when neglecting the effects of the kurtosis, etc., in the calculation of the halo mass function.}

A possible solution to this problem is to take a $\fNL(k)$ such that it is constant up to some scale $k_\text{\tiny cut}$ (corresponding to an halo mass $M_\text{\tiny cut}$) and vanishing at smaller momenta, that is (see e.g. \cite{Sabti:2020ser})
\begin{align}
{\it ii)}\qquad
B_\zeta(k_1,k_2,k_3) = \frac{6}{5} 
\fNL^0 P_\zeta(k_2)P_\zeta(k_3) 
\left[\prod_{i=1}^3
\Theta(k_i - k_\text{\tiny cut})
+ {\rm perm.}\right].
\end{align}
Such a scale dependent NG can be obtained in inflationary models where besides the inflaton field there is another spectator field which experiences a transition from massless to massive at a scale $\approx k_\text{\tiny cut}$ \cite{Riotto:2010nh}.
The corresponding result is shown in Fig.~\ref{fig:Labberunning} (central panel), where
we plot the stellar mass density above $M_*$ assuming different values of $\fNL^0$ 
and the scale where NGs are switched off corresponding to
$M_\text{\tiny cut}= 10^{12} M_\odot$.
This indicates that, in order to reach the JWST observations, large NGs are needed at least starting from masses below $\approx 10^{12} M_\odot$.
The resulting shape of $S_3$ obtained in this scenario is shown in Fig.~\ref{fig:S3running}.
 Note that these large values of $\fNL^0$ might be in tension with constraints quoted by \cite{Sabti:2020ser}, which strongly depend on  $k_{\rm cut}$.

Finally, we consider a model in which the NG correction is localised within a bump at scales close the one observed by Ref.~\cite{2022arXiv220712446L}.
We assume the functional form 
\begin{equation}
{\it iii)}\qquad
\fNL(k) = 
\frac{\fNL^0 }{\sqrt{2 \pi} w}
\exp\llp -\frac{\log^2 \lp k/k_0\rp}{2 w^2}\rrp .
\end{equation}
and fix the central scale  to be $k_0 = 1.4 h/{\rm Mpc}$
which corresponds to the masses detected by \cite{2022arXiv220712446L} at redshift $z=10$. 
In Fig.~\ref{fig:Labberunning} we show the corresponding stellar mass density above $M_*$ with varying assumptions on 
the width of the bump $w$, while the resulting skewness is plotted in Fig.~\ref{fig:S3running}.
We see that a large normalisation $\fNL^0$
and a relatively narrow width may allow to reduce the tension between JWST observations and the cosmological model.

%%%%%%%%%%%%%%%%%%%%%%%%%%%%%%%%%%%%%%%%%%%%%%%%%%%%%%
\section{Conclusions}\label{sec:conclusions}
%%%%%%%%%%%%%%%%%%%%%%%%%%%%%%%%%%%%%%%%%%%%%%%%%%%%%%
\noindent
In this paper we have investigated whether changing the initial conditions of the cosmological perturbations by adding some amount of NG helps in boosting the formation of massive and bright galaxies, as recently reported in the literature thanks to the new data collected by the JWST. 

We tested our modelling of the NG correction of the halo mass function adopting N-body simulations and check whether NG scenarios compatible with current large-scale and low redshift observations may help explaining recent data. 
Our findings indicate that a large and strongly scale dependent NG (which switches on at small scales) is needed to alleviate the tension between the cosmological model and the observations.

We have modelled the halo mass distribution with the Tinker model \cite{Tinker:2010my}, which was used in the model validation against N-body simulations.
Notice, however, that different choices were adopted in the literature (e.g. Sheth-Tormen \cite{Sheth:2001dp}) that lead to larger HMF tail and, consequently, to smaller values of $\fNL$ to alleviate the tension. We have verified this intuition by performing our analysis using the Sheth-Tormen mass function, for which the values of $\fNL$ needed to alleviate the tension are around a factor of two smaller.

We notice once again that the small evolution of the halo mass function between redshift $8$ and $10$ reported in Ref.~\cite{2022arXiv220712446L} poses a threat to our explanation as it is not easily captured in the models we tested and would need a rather artificial redshift dependence of the theoretical prediction. Such a caveat appears to be valid for most of the solutions recently proposed in the literature.
It should be noted that our analysis does not include a complete assessment of parameter degeneracies within the $\Lambda$CDM model. In particular, $\sigma_8$, which parametrizes the amplitude of matter fluctuations, is known to also provide an enhancement on the tail of the HMF. Leaving all other cosmological parameters fixed and setting $\fNL=0$, we have verified that to explain the observed galaxies would require values of $\sigma_8\gtrsim 0.9$ which are significantly excluded by Planck \cite{Planck:2018vyg}.

We are aware that there are several uncertainties related to the JWST measurements that might solve the tension with respect to the $\Lambda$CDM independently from NG.  
First of all, uncertainties in the calibration of JWST data may impact the redshift determinations, see e.g. table 4 in \cite{2023MNRAS.518.4755A}.
Current measurements rely on identifying high redshift candidates using photometric template fitting  which however  are not tested at such high redshifts \cite{2022arXiv220807879S}. 
Another  possibly large  uncertainty is added in the estimation of $M_*$. 
For it the  Chabrier Initial Mass Function is typically adopted  \cite{Chabrier:2003ki}, which however is tested at much lower masses and redshifts. Furthermore, the effect of a large scatter in the star formation  \cite{2022arXiv220812826M} as well as the impact of dust attenuation \cite{2022arXiv220906840Z} may introduce further contamination in the mass estimation. 
Finally, if the data also include very bright lines from other sources (such as  AGN) beyond the stellar continuum, one may also get a contaminated 
 measurement of masses (see e.g. \cite{2022arXiv220814999E}).
The spectroscopic follow-up and further testing on the astrophysical uncertainties will soon shed more light on the issue.

%\begin{acknowledgments}
\vspace{.7 cm}
\noindent
The authors would like to thank Pierluigi Monaco for very useful  comments on the draft and for discussions on the possible sources of uncertainty on the JWST measurements.   M.B. acknowledges support from the NWO project “Cosmic origins from simulated universes” for the computing time allocated to run a subset of the Eos simulations
on \textsc{Snellius}, a supercomputer that is part of the Dutch National Computing Facilities.
G.~F.  acknowledges financial support provided under the European
Union's H2020 ERC, Starting Grant agreement no.~DarkGRA--757480 and under the MIUR PRIN programme, and support from the Amaldi Research Center funded by the MIUR program ``Dipartimento di Eccellenza" (CUP:~B81I18001170001). This work was supported by the EU Horizon 2020 Research and Innovation Programme under the Marie Sklodowska-Curie Grant Agreement No. 101007855. A.R. acknowledges financial support provided by the Boninchi Foundation. 

%\end{acknowledgments}

\bibliographystyle{aasjournal}
\bibliography{main}

\begin{thebibliography}{}
\expandafter\ifx\csname natexlab\endcsname\relax\def\natexlab#1{#1}\fi
\providecommand{\url}[1]{\href{#1}{#1}}
\providecommand{\dodoi}[1]{doi:~\href{http://doi.org/#1}{\nolinkurl{#1}}}
\providecommand{\doeprint}[1]{\href{http://ascl.net/#1}{\nolinkurl{http://ascl.net/#1}}}
\providecommand{\doarXiv}[1]{\href{https://arxiv.org/abs/#1}{\nolinkurl{https://arxiv.org/abs/#1}}}

\bibitem[{Achitouv {et~al.}(2014)Achitouv, Wagner, Weller, \&
  Rasera}]{Achitouv:2013oea}
Achitouv, I., Wagner, C., Weller, J., \& Rasera, Y. 2014, JCAP, 10, 077,
  \dodoi{10.1088/1475-7516/2014/10/077}

\bibitem[{Achitouv \& Corasaniti(2012)}]{Achitouv:2011sq}
Achitouv, I.~E., \& Corasaniti, P.~S. 2012, JCAP, 1202, 002,
  \dodoi{10.1088/1475-7516/2012/07/E01, 10.1088/1475-7516/2012/02/002}

\bibitem[{{Adams} {et~al.}(2023){Adams}, {Conselice}, {Ferreira}, {Austin},
  {Trussler}, {Juod{\v{z}}balis}, {Wilkins}, {Caruana}, {Dayal}, {Verma}, \&
  {Vijayan}}]{2023MNRAS.518.4755A}
{Adams}, N.~J., {Conselice}, C.~J., {Ferreira}, L., {et~al.} 2023, \mnras, 518,
  4755, \dodoi{10.1093/mnras/stac3347}

\bibitem[{Aghanim {et~al.}(2020)}]{Planck:2018vyg}
Aghanim, N., {et~al.} 2020, Astron. Astrophys., 641, A6,
  \dodoi{10.1051/0004-6361/201833910}

\bibitem[{Agullo \& Shandera(2012)}]{Agullo:2012cs}
Agullo, I., \& Shandera, S. 2012, JCAP, 09, 007,
  \dodoi{10.1088/1475-7516/2012/09/007}

\bibitem[{Akrami {et~al.}(2020)}]{Planck:2019kim}
Akrami, Y., {et~al.} 2020, Astron. Astrophys., 641, A9,
  \dodoi{10.1051/0004-6361/201935891}

\bibitem[{{Atek} {et~al.}(2022){Atek}, {Shuntov}, {Furtak}, {Richard}, {Kneib},
  {Mahler Adi Zitrin}, {McCracken}, {Laigle}, \&
  {Charlot}}]{2022arXiv220712338A}
{Atek}, H., {Shuntov}, M., {Furtak}, L.~J., {et~al.} 2022, arXiv e-prints,
  arXiv:2207.12338.
\newblock \doarXiv{2207.12338}

\bibitem[{Bartolo {et~al.}(2004)Bartolo, Komatsu, Matarrese, \&
  Riotto}]{Bartolo:2004if}
Bartolo, N., Komatsu, E., Matarrese, S., \& Riotto, A. 2004, Phys. Rept., 402,
  103, \dodoi{10.1016/j.physrep.2004.08.022}

\bibitem[{Becker {et~al.}(2011)Becker, Huterer, \& Kadota}]{Becker:2010hx}
Becker, A., Huterer, D., \& Kadota, K. 2011, JCAP, 01, 006,
  \dodoi{10.1088/1475-7516/2011/01/006}

\bibitem[{Becker {et~al.}(2012)Becker, Huterer, \& Kadota}]{Becker:2012yr}
---. 2012, JCAP, 12, 034, \dodoi{10.1088/1475-7516/2012/12/034}

\bibitem[{Behroozi {et~al.}(2013)Behroozi, Wechsler, \& Wu}]{Behroozi:2011ju}
Behroozi, P.~S., Wechsler, R.~H., \& Wu, H.-Y. 2013, Astrophys. J., 762, 109,
  \dodoi{10.1088/0004-637X/762/2/109}

\bibitem[{Biagetti(2019)}]{Biagetti:2019bnp}
Biagetti, M. 2019, Galaxies, 7, 71, \dodoi{10.3390/galaxies7030071}

\bibitem[{Biagetti {et~al.}(2017)Biagetti, Lazeyras, Baldauf, Desjacques, \&
  Schmidt}]{Biagetti:2016ywx}
Biagetti, M., Lazeyras, T., Baldauf, T., Desjacques, V., \& Schmidt, F. 2017,
  Mon. Not. Roy. Astron. Soc., 468, 3277, \dodoi{10.1093/mnras/stx714}

\bibitem[{Biagetti {et~al.}(2013)Biagetti, Perrier, Riotto, \&
  Desjacques}]{Biagetti:2013sr}
Biagetti, M., Perrier, H., Riotto, A., \& Desjacques, V. 2013, Phys. Rev. D,
  87, 063521, \dodoi{10.1103/PhysRevD.87.063521}

\bibitem[{Blas {et~al.}(2011)Blas, Lesgourgues, \& Tram}]{Blas:2011rf}
Blas, D., Lesgourgues, J., \& Tram, T. 2011, JCAP, 1107, 034,
  \dodoi{10.1088/1475-7516/2011/07/034}

\bibitem[{Borgani {et~al.}(1994)Borgani, Coles, Moscardini, \&
  Plionis}]{Borgani:1993nz}
Borgani, S., Coles, P., Moscardini, L., \& Plionis, M. 1994, Mon. Not. Roy.
  Astron. Soc., 266, 524, \dodoi{10.1093/mnras/266.2.524}

\bibitem[{Byrnes {et~al.}(2011)Byrnes, Enqvist, Nurmi, \&
  Takahashi}]{Byrnes:2011gh}
Byrnes, C.~T., Enqvist, K., Nurmi, S., \& Takahashi, T. 2011, JCAP, 11, 011,
  \dodoi{10.1088/1475-7516/2011/11/011}

\bibitem[{Byrnes {et~al.}(2010)Byrnes, Gerstenlauer, Nurmi, Tasinato, \&
  Wands}]{Byrnes:2010ft}
Byrnes, C.~T., Gerstenlauer, M., Nurmi, S., Tasinato, G., \& Wands, D. 2010,
  JCAP, 10, 004, \dodoi{10.1088/1475-7516/2010/10/004}

\bibitem[{Castorina {et~al.}(2019)}]{Castorina:2019wmr}
Castorina, E., {et~al.} 2019.
\newblock \doarXiv{1904.08859}

\bibitem[{Chabrier(2003)}]{Chabrier:2003ki}
Chabrier, G. 2003, Publ. Astron. Soc. Pac., 115, 763, \dodoi{10.1086/376392}

\bibitem[{Chen(2005)}]{Chen:2005fe}
Chen, X. 2005, Phys. Rev. D, 72, 123518, \dodoi{10.1103/PhysRevD.72.123518}

\bibitem[{Chongchitnan \& Silk(2010)}]{Chongchitnan:2010xz}
Chongchitnan, S., \& Silk, J. 2010, Astrophys. J., 724, 285,
  \dodoi{10.1088/0004-637X/724/1/285}

\bibitem[{Cohn \& White(2008)}]{Cohn:2007xu}
Cohn, J.~D., \& White, M.~J. 2008, Mon. Not. Roy. Astron. Soc., 385, 2025,
  \dodoi{10.1111/j.1365-2966.2008.12972.x}

\bibitem[{Dalal {et~al.}(2008)Dalal, Dor\'e, Huterer, \&
  Shirokov}]{Dalal:2007cu}
Dalal, N., Dor\'e, O., Huterer, D., \& Shirokov, A. 2008, Phys. Rev., D77,
  123514, \dodoi{10.1103/PhysRevD.77.123514}

\bibitem[{Desjacques \& Seljak(2010)}]{Desjacques:2009jb}
Desjacques, V., \& Seljak, U. 2010, Phys. Rev., D81, 023006,
  \dodoi{10.1103/PhysRevD.81.023006}

\bibitem[{Despali {et~al.}(2015)Despali, Giocoli, Angulo, Tormen, Sheth, Baso,
  \& Moscardini}]{Despali:2015yla}
Despali, G., Giocoli, C., Angulo, R.~E., {et~al.} 2015,
  \dodoi{10.1093/mnras/stv2842}

\bibitem[{{Endsley} {et~al.}(2022){Endsley}, {Stark}, {Whitler}, {Topping},
  {Chen}, {Plat}, {Chisholm}, \& {Charlot}}]{2022arXiv220814999E}
{Endsley}, R., {Stark}, D.~P., {Whitler}, L., {et~al.} 2022, arXiv e-prints,
  arXiv:2208.14999.
\newblock \doarXiv{2208.14999}

\bibitem[{Evrard {et~al.}(2002)}]{VIRGO:2001szp}
Evrard, A.~E., {et~al.} 2002, Astrophys. J., 573, 7, \dodoi{10.1086/340551}

\bibitem[{{Finkelstein} {et~al.}(2022){Finkelstein}, {Bagley}, {Arrabal Haro},
  {Dickinson}, {Ferguson}, {Kartaltepe}, {Papovich}, {Burgarella}, {Kocevski},
  {Huertas-Company}, {Iyer}, {Larson}, {P{\'e}rez-Gonz{\'a}lez}, {Rose},
  {Tacchella}, {Wilkins}, {Chworowsky}, {Medrano}, {Morales}, {Somerville},
  {Yung}, {Fontana}, {Giavalisco}, {Grazian}, {Grogin}, {Kewley}, {Koekemoer},
  {Kirkpatrick}, {Kurczynski}, {Lotz}, {Pentericci}, {Pirzkal}, {Ravindranath},
  {Ryan}, {Trump}, {Yang}, {Almaini}, {Amor{\'\i}n}, {Annunziatella},
  {Backhaus}, {Barro}, {Behroozi}, {Bell}, {Bhatawdekar}, {Bisigello}, {Bromm},
  {Buat}, {Buitrago}, {Calabr{\'o}}, {Casey}, {Castellano}, {Ch{\'a}vez Ortiz},
  {Ciesla}, {Cleri}, {Cohen}, {Cole}, {Cooke}, {Cooper}, {Cooray}, {Costantin},
  {Cox}, {Croton}, {Daddi}, {Dav{\'e}}, {de la Vega}, {Dekel}, {Elbaz},
  {Estrada-Carpenter}, {Faber}, {Fern{\'a}ndez}, {Finkelstein}, {Freundlich},
  {Fujimoto}, {Garc{\'\i}a-Argum{\'a}nez}, {Gardner}, {Gawiser},
  {G{\'o}mez-Guijarro}, {Guo}, {Hamilton}, {Hathi}, {Holwerda}, {Hirschmann},
  {Hutchison}, {Jaskot}, {Jha}, {Jogee}, {Juneau}, {Jung}, {Kassin}, {Le Bail},
  {Leung}, {Lucas}, {Magnelli}, {Mantha}, {Matharu}, {McGrath}, {McIntosh},
  {Merlin}, {Mobasher}, {Newman}, {Nicholls}, {Pandya}, {Rafelski}, {Ronayne},
  {Santini}, {Seill{\'e}}, {Shah}, {Shen}, {Simons}, {Snyder}, {Stanway},
  {Straughn}, {Teplitz}, {Vanderhoof}, {Vega-Ferrero}, {Wang}, {Weiner},
  {Willmer}, {Wuyts}, \& {Zavala}}]{2022arXiv220712474F}
{Finkelstein}, S.~L., {Bagley}, M.~B., {Arrabal Haro}, P., {et~al.} 2022, arXiv
  e-prints, arXiv:2207.12474.
\newblock \doarXiv{2207.12474}

\bibitem[{Giannantonio {et~al.}(2012)Giannantonio, Porciani, Carron, Amara, \&
  Pillepich}]{Giannantonio:2011ya}
Giannantonio, T., Porciani, C., Carron, J., Amara, A., \& Pillepich, A. 2012,
  Mon. Not. Roy. Astron. Soc., 422, 2854,
  \dodoi{10.1111/j.1365-2966.2012.20604.x}

\bibitem[{Gong {et~al.}(2022)Gong, Yue, Cao, \& Chen}]{Gong:2022qjx}
Gong, Y., Yue, B., Cao, Y., \& Chen, X. 2022.
\newblock \doarXiv{2209.13757}

\bibitem[{{Gooding} {et~al.}(1992){Gooding}, {Park}, {Spergel}, {Turok}, \&
  {Gott}}]{Gooding:1991ys}
{Gooding}, A.~K., {Park}, C., {Spergel}, D.~N., {Turok}, N., \& {Gott},
  Richard, I. 1992, \apj, 393, 42, \dodoi{10.1086/171483}

\bibitem[{{Harikane} {et~al.}(2022){Harikane}, {Ouchi}, {Oguri}, {Ono},
  {Nakajima}, {Isobe}, {Umeda}, {Mawatari}, \& {Zhang}}]{2022arXiv220801612H}
{Harikane}, Y., {Ouchi}, M., {Oguri}, M., {et~al.} 2022, arXiv e-prints,
  arXiv:2208.01612.
\newblock \doarXiv{2208.01612}

\bibitem[{Huang(2010{\natexlab{a}})}]{Huang:2010cy}
Huang, Q.-G. 2010{\natexlab{a}}, JCAP, 11, 026,
  \dodoi{10.1088/1475-7516/2011/02/E01}

\bibitem[{Huang(2010{\natexlab{b}})}]{Huang:2010es}
---. 2010{\natexlab{b}}, JCAP, 12, 017, \dodoi{10.1088/1475-7516/2010/12/017}

\bibitem[{Jenkins {et~al.}(2001)Jenkins, Frenk, White, Colberg, Cole, Evrard,
  Couchman, \& Yoshida}]{Jenkins:2000bv}
Jenkins, A., Frenk, C.~S., White, S. D.~M., {et~al.} 2001, Mon. Not. Roy.
  Astron. Soc., 321, 372, \dodoi{10.1046/j.1365-8711.2001.04029.x}

\bibitem[{Khoury \& Piazza(2009)}]{Khoury:2008wj}
Khoury, J., \& Piazza, F. 2009, JCAP, 07, 026,
  \dodoi{10.1088/1475-7516/2009/07/026}

\bibitem[{{Labbe} {et~al.}(2022){Labbe}, {van Dokkum}, {Nelson}, {Bezanson},
  {Suess}, {Leja}, {Brammer}, {Whitaker}, {Mathews}, \&
  {Stefanon}}]{2022arXiv220712446L}
{Labbe}, I., {van Dokkum}, P., {Nelson}, E., {et~al.} 2022, arXiv e-prints,
  arXiv:2207.12446.
\newblock \doarXiv{2207.12446}

\bibitem[{Liu \& Bromm(2022)}]{Liu:2022bvr}
Liu, B., \& Bromm, V. 2022.
\newblock \doarXiv{2208.13178}

\bibitem[{Lovell {et~al.}(2022)Lovell, Harrison, Harikane, Tacchella, \&
  Wilkins}]{Lovell:2022bhx}
Lovell, C.~C., Harrison, I., Harikane, Y., Tacchella, S., \& Wilkins, S.~M.
  2022.
\newblock \doarXiv{2208.10479}

\bibitem[{LoVerde {et~al.}(2008)LoVerde, Miller, Shandera, \&
  Verde}]{LoVerde:2007ri}
LoVerde, M., Miller, A., Shandera, S., \& Verde, L. 2008, JCAP, 04, 014,
  \dodoi{10.1088/1475-7516/2008/04/014}

\bibitem[{Lukic {et~al.}(2007)Lukic, Heitmann, Habib, Bashinsky, \&
  Ricker}]{Lukic:2007fc}
Lukic, Z., Heitmann, K., Habib, S., Bashinsky, S., \& Ricker, P.~M. 2007,
  Astrophys. J., 671, 1160, \dodoi{10.1086/523083}

\bibitem[{Matarrese {et~al.}(1991)Matarrese, Lucchin, Messina, \&
  Moscardini}]{Matarrese:1991sj}
Matarrese, S., Lucchin, F., Messina, A., \& Moscardini, L. 1991, Mon. Not. Roy.
  Astron. Soc., 253, 35

\bibitem[{Matarrese {et~al.}(2000)Matarrese, Verde, \&
  Jimenez}]{Matarrese:2000iz}
Matarrese, S., Verde, L., \& Jimenez, R. 2000, Astrophys. J., 541, 10,
  \dodoi{10.1086/309412}

\bibitem[{Menci {et~al.}(2022)Menci, Castellano, Santini, Merlin, Fontana, \&
  Shankar}]{Menci:2022wia}
Menci, N., Castellano, M., Santini, P., {et~al.} 2022.
\newblock \doarXiv{2208.11471}

\bibitem[{{Mirocha} \& {Furlanetto}(2022)}]{2022arXiv220812826M}
{Mirocha}, J., \& {Furlanetto}, S.~R. 2022, arXiv e-prints, arXiv:2208.12826.
\newblock \doarXiv{2208.12826}

\bibitem[{Moscardini {et~al.}(1991)Moscardini, Matarrese, Lucchin, \&
  Messina}]{Moscardini:1990zh}
Moscardini, L., Matarrese, S., Lucchin, F., \& Messina, A. 1991, Mon. Not. Roy.
  Astron. Soc., 248, 424

\bibitem[{{Naidu} {et~al.}(2022){Naidu}, {Oesch}, {Setton}, {Matthee},
  {Conroy}, {Johnson}, {Weaver}, {Bouwens}, {Brammer}, {Dayal}, {Illingworth},
  {Barrufet}, {Belli}, {Bezanson}, {Bose}, {Heintz}, {Leja}, {Leonova},
  {Marques-Chaves}, {Stefanon}, {Toft}, {van der Wel}, {van Dokkum}, {Weibel},
  \& {Whitaker}}]{2022arXiv220802794N}
{Naidu}, R.~P., {Oesch}, P.~A., {Setton}, D.~J., {et~al.} 2022, arXiv e-prints,
  arXiv:2208.02794.
\newblock \doarXiv{2208.02794}

\bibitem[{{Park} {et~al.}(1991){Park}, {Spergel}, \& {Turok}}]{Park:1991mh}
{Park}, C., {Spergel}, D.~N., \& {Turok}, N. 1991, \apjl, 372, L53,
  \dodoi{10.1086/186022}

\bibitem[{Pillepich {et~al.}(2010)Pillepich, Porciani, \&
  Hahn}]{Pillepich:2008ka}
Pillepich, A., Porciani, C., \& Hahn, O. 2010, Mon. Not. Roy. Astron. Soc.,
  402, 191, \dodoi{10.1111/j.1365-2966.2009.15914.x}

\bibitem[{Reed {et~al.}(2007)Reed, Bower, Frenk, Jenkins, \&
  Theuns}]{Reed:2006rw}
Reed, D., Bower, R., Frenk, C., Jenkins, A., \& Theuns, T. 2007, Mon. Not. Roy.
  Astron. Soc., 374, 2, \dodoi{10.1111/j.1365-2966.2006.11204.x}

\bibitem[{Reed {et~al.}(2003)Reed, Gardner, Quinn, Stadel, Fardal, Lake, \&
  Governato}]{Reed:2003sq}
Reed, D., Gardner, J., Quinn, T.~R., {et~al.} 2003, Mon. Not. Roy. Astron.
  Soc., 346, 565, \dodoi{10.1046/j.1365-2966.2003.07113.x}

\bibitem[{Riotto \& Sloth(2011)}]{Riotto:2010nh}
Riotto, A., \& Sloth, M.~S. 2011, Phys. Rev. D, 83, 041301,
  \dodoi{10.1103/PhysRevD.83.041301}

\bibitem[{Sabti {et~al.}(2021)Sabti, Mu\~noz, \& Blas}]{Sabti:2020ser}
Sabti, N., Mu\~noz, J.~B., \& Blas, D. 2021, JCAP, 01, 010,
  \dodoi{10.1088/1475-7516/2021/01/010}

\bibitem[{Scoccimarro(1998)}]{Scoccimarro:1997gr}
Scoccimarro, R. 1998, Mon. Not. Roy. Astron. Soc., 299, 1097,
  \dodoi{10.1046/j.1365-8711.1998.01845.x}

\bibitem[{Scoccimarro {et~al.}(2012)Scoccimarro, Hui, Manera, \&
  Chan}]{Scoccimarro:2011pz}
Scoccimarro, R., Hui, L., Manera, M., \& Chan, K.~C. 2012, Phys. Rev., D85,
  083002, \dodoi{10.1103/PhysRevD.85.083002}

\bibitem[{Sefusatti {et~al.}(2009)Sefusatti, Liguori, Yadav, Jackson, \&
  Pajer}]{Sefusatti:2009xu}
Sefusatti, E., Liguori, M., Yadav, A. P.~S., Jackson, M.~G., \& Pajer, E. 2009,
  JCAP, 12, 022, \dodoi{10.1088/1475-7516/2009/12/022}

\bibitem[{Sheth \& Tormen(2002)}]{Sheth:2001dp}
Sheth, R.~K., \& Tormen, G. 2002, Mon. Not. Roy. Astron. Soc., 329, 61,
  \dodoi{10.1046/j.1365-8711.2002.04950.x}

\bibitem[{Springel(2005)}]{Springel:2005mi}
Springel, V. 2005, Mon. Not. Roy. Astron. Soc., 364, 1105,
  \dodoi{10.1111/j.1365-2966.2005.09655.x}

\bibitem[{Stahl {et~al.}(2022)Stahl, Montandon, Famaey, Hahn, \&
  Ibata}]{Stahl:2022did}
Stahl, C., Montandon, T., Famaey, B., Hahn, O., \& Ibata, R. 2022.
\newblock \doarXiv{2209.15038}

\bibitem[{{Steinhardt} {et~al.}(2022){Steinhardt}, {Kokorev}, {Rusakov},
  {Garcia}, \& {Sneppen}}]{2022arXiv220807879S}
{Steinhardt}, C.~L., {Kokorev}, V., {Rusakov}, V., {Garcia}, E., \& {Sneppen},
  A. 2022, arXiv e-prints, arXiv:2208.07879.
\newblock \doarXiv{2208.07879}

\bibitem[{Tinker {et~al.}(2008)Tinker, Kravtsov, Klypin, Abazajian, Warren,
  Yepes, Gottlober, \& Holz}]{Tinker:2008ff}
Tinker, J.~L., Kravtsov, A.~V., Klypin, A., {et~al.} 2008, Astrophys. J., 688,
  709, \dodoi{10.1086/591439}

\bibitem[{Tinker {et~al.}(2010)Tinker, Robertson, Kravtsov, Klypin, Warren,
  Yepes, \& Gottlober}]{Tinker:2010my}
Tinker, J.~L., Robertson, B.~E., Kravtsov, A.~V., {et~al.} 2010, Astrophys. J.,
  724, 878, \dodoi{10.1088/0004-637X/724/2/878}

\bibitem[{Warren {et~al.}(2006)Warren, Abazajian, Holz, \&
  Teodoro}]{Warren:2005ey}
Warren, M.~S., Abazajian, K., Holz, D.~E., \& Teodoro, L. 2006, Astrophys. J.,
  646, 881, \dodoi{10.1086/504962}

\bibitem[{Weinberg \& Cole(1992)}]{Weinberg:1991qe}
Weinberg, D.~H., \& Cole, S. 1992, Mon. Not. Roy. Astron. Soc., 259, 652

\bibitem[{Yan {et~al.}(2022)Yan, Ma, Ling, Cheng, Huang, \&
  Zitrin}]{Yan:2022sxd}
Yan, H., Ma, Z., Ling, C., {et~al.} 2022.
\newblock \doarXiv{2207.11558}

\bibitem[{{Ziparo} {et~al.}(2022){Ziparo}, {Ferrara}, {Sommovigo}, \&
  {Kohandel}}]{2022arXiv220906840Z}
{Ziparo}, F., {Ferrara}, A., {Sommovigo}, L., \& {Kohandel}, M. 2022, arXiv
  e-prints, arXiv:2209.06840.
\newblock \doarXiv{2209.06840}

\end{thebibliography}

\end{document}